\documentclass[prl,twocolumn,superscriptaddress,floatfix, preprintnumbers]{revtex4-1} 
\usepackage{epsfig}
\usepackage{graphicx}
\usepackage{palatino}
\usepackage{hyphenat}
\usepackage{amsmath}
\usepackage{amssymb}
\usepackage{mathtools}
\usepackage{mathrsfs}
\usepackage{slashed}
\usepackage{epstopdf}
\usepackage{xcolor}
\usepackage{braket}
\usepackage{booktabs}
\definecolor{lcolor}{rgb}{0.,0.0,0.}
\definecolor{citcolor}{rgb}{0,0.,0.5}
\usepackage[breaklinks,colorlinks,urlcolor=blue,citecolor=blue,linkcolor=blue]{hyperref}
\usepackage{multirow}
\usepackage{ltablex}
\usepackage{soul}

\newcommand{\beq}{\begin{eqnarray}}
\newcommand{\eeq}{\end{eqnarray}}

\newcommand{\bem}{\begin{multline}}
\newcommand{\eem}{\end{multline}}
\newcommand{\beg}{\begin{gather}}
\newcommand{\eeg}{\end{gather}}

\newcommand{\nn}{\nonumber\\}

\newcommand{\ben}{\begin{eqnarray*}}
\newcommand{\een}{\end{eqnarray*}}

\newcommand{\secn}[1]{Section~1}
\newcommand{\appn}[1]{Appendix~1}

\long\def\comment#1{ }

\def\and{\quad\text{and}\quad}

\def\ua{{ \uparrow}}
\def\da{{ \downarrow}}

\def\0{{\boldsymbol 0}}

\def\0{{\boldsymbol 0}}

\def\Hs{H_{\rm Schwinger }}

\begin{document}
\preprint{MIT-CTP/5599}

\title{Realtime dynamics of hyperon spin correlations\\ from string fragmentation in a deformed four-flavor Schwinger model}

\author{João Barata}
\email[]{jlourenco@bnl.gov}
\affiliation{Physics Department, Brookhaven National Laboratory, Upton, NY 11973, USA}
\author{Wenjie Gong}
\email[]{wgong@mit.edu}
\affiliation{Center for Theoretical Physics, Massachusetts Institute of Technology, Cambridge, MA 02139, USA}
\author{Raju Venugopalan}
\email[]{raju@bnl.gov}
\affiliation{Physics Department, Brookhaven National Laboratory, Upton, NY 11973, USA}
\affiliation{Center for Frontiers in Nuclear Science, Stony Brook University, Stony Brook, NY 11794}

\begin{abstract}
Self-polarizing weak decays of $\Lambda$-hyperons  provide unique insight into the role of entanglement in the fragmentation of QCD strings through measurements of the spin correlations of $\Lambda{\bar \Lambda}$-pairs produced in collider experiments. 
The simplest quantum field theory representing the underlying parton dynamics is the four-flavor massive Schwinger model 
plus an effective spin-flip term, where the flavors are mapped to light (up/down) and heavy (strange) quarks and their spins. This construction provides a novel way to explore hyperon spin-correlations 
in 1+1-dimensions. 
We investigate the evolution of these correlations for different string configurations that are sensitive to the rich structure of the model Hamiltonian. 

\end{abstract}

\maketitle

An intriguing question in QCD and QCD-like theories is the role of entanglement in the confinement of quarks and gluons (partons) within hadrons~\cite{Klebanov:2007ws,Beane:2018oxh}. In high energy QCD, features of entanglement 
were explored recently in the context of parton distributions in deeply inelastic scattering~\cite{Kharzeev:2017qzs,Armesto:2019mna,
Hentschinski:2021aux,Dumitru:2022tud,Dumitru:2023qee,Hentschinski:2023izh,Asadi:2022vbl,Duan:2023zls,Ramos:2020kaj,Beane:2019loz} and in parton fragmentation into hadrons~\cite{Berges:2017zws,Berges:2017hne,Neill:2018uqw,
Gong:2021bcp,Benito-Calvino:2022kqa,Barata:2023uoi,Barata:2023clv}.  A strong motivation is the promise of fresh insight from  quantum information science (QIS) into these fundamental quantum many-body parton features of hadrons ~\cite{Dvali:2021ooc,Dvali:2021jto,Kharzeev:2021nzh,Kou:2022dkw,Klco:2021biu}. 

 Several proposals have emerged for quantum entanglement and Bell-type inequality measures in the challenging environment of collider experiments~\cite{Hao:2009kj,Barr:2021zcp,Fabbrichesi:2021npl,Afik:2022dgh,Klco:2023ojt,Ashby-Pickering:2022umy,Florio:2023dke}. 
 In  \cite{Gong:2021bcp}, two of us proposed that the self-polarizing weak decays of hyperons can be exploited to measure  $\Lambda {\bar \Lambda}$ spin correlations in the fragmentation of QCD strings. Following earlier work~\cite{Tornqvist:1980af} on the Clauser-Horne-Shimony-Holt (CHSH) inequality~\cite{PhysRevLett.23.880} for $\Lambda {\bar\Lambda}$ spin correlations, we constructed a modified CHSH inequality and entanglement measures of the string spin density matrix. Since $\Lambda$ (${\bar \Lambda}$) hyperons 
 contain a flavor triplet of up, down and strange (anti)quarks, 
 measurements of their spin correlations
 probe quantum features of parton dynamics within QCD strings. In particular, clean 
 extraction of hyperon spin correlations~\cite{Tu:2023few} will be possible at the future Electron Ion Collider~\cite{Accardi:2012qut}. 
 
 Further progress exploiting the potential of hyperon spin correlations requires a dynamical model of parton dynamics in the formation and fragmentation of QCD strings. Nonperturbative first principles  quantum field theory (QFT) methods such as lattice QCD are inapplicable 
 because of the intrinsically realtime dynamics of hadronization. 
 Therefore, phenomenological models are the state-of-the art in describing hadronization at colliders,  
 the quintessential example being the Lund string model~\cite{AnderssonPR1983} implemented in the widely used event generator PYTHIA~\cite{Sjostrand:2014zea}. Since this model is a semi-classical implementation of 1+1-d QED (also known as the Schwinger model), computations on quantum devices offer a promising path towards performing first principles simulations of string dynamics~\cite{Banuls:2019bmf,DiMeglio:2023nsa,Bauer:2023qgm} that may be relevant to understanding quantum features of hadronization at colliders. In particular, such studies can help towards implementing QIS features in phenomenological semi-classical frameworks in the near future~\cite{Hunt-Smith:2020lul}.

 In this paper, we will outline the simplest 1+1-dimensional QFT that captures the rich flavor and spin dynamics of light and heavy partons in the QCD string, allowing us to model the real-time production and evolution of light and heavy quark flavors in a QCD string for the first time. It significantly extends the 
 static string configuration study in~\cite{Gong:2021bcp}. Further, our QFT  provides novel insight that can help identify quantum 
 features of measured $\Lambda {\bar \Lambda}$-correlations; this effort is   timely, motivating experimental measurements of hyperon spin correlations at colliders~\cite{Vanek:2023oeo}.

 We will assume, as in the nonrelativistic quark model, that $\Lambda$-spin is carried by the heavy strange quark. The status of this ansatz is uncertain; for proposed experimental tests, see \cite{Burkardt:1993zh,Moretti:2019peg,Ellis:2011kq,Metz:2016swz,Tu:2023few}. An interesting, albeit nontrivial, extension of our study would be to construct baryons in a similar 1+1-dimensional setup and contrast their "spin"-correlations with those of their strange quark constituents. In this present work, we model hyperon spin-correlations by solely studying heavy-quark spin correlations.
 
 To construct our QFT model of parton dynamics in the QCD string, we start from the massive Schwinger model with $N_f=4$ quark flavors~\footnote{The massive $N_f > 1$ Schwinger model is qualitatively different from the $N_f=1$ case~\cite{Coleman:1976uz,Steinhardt77,Hetrick:1995wq,Delphenich:1997ex}. For $N_f > 2$ generalizations, see~\cite{Berruto:1998tg,Hosotani:1998kd,Berruto:1999ga,Hetrick:1995wq,Delphenich:1997ex}.}, for reasons we will elaborate on shortly.
In temporal gauge $A_0=0$, this model  is described by the Hamiltonian
\begin{align}\label{eq:H_Schwinger}
	&\Hs=  \int dx \, \frac{1}{2 } E^2(x) \nn 
    +&  \sum_{f=1}^{N_f=4} \bar  \psi_f(x) (-i \gamma^1 \partial_1 +g \gamma^1 A_1(x) +m_f ) \psi_f(x)\, ,
\end{align}
where $g$ is the dimensionful coupling constant and the electric field $E = -F^{01} = -\partial^0 A^1$, with  $A_1$ denoting the other component of the $U(1)$ gauge field. Note that there is no magnetic field in 1+1-dimensions. The two component fermion spinors $\psi_f$, with flavor indices $f=1,2,3,4$, satisfy the canonical commutation relations, 
\begin{align}\label{eq:com_rels}
&\{\psi_{f,\alpha}(x),\psi_{l,\beta}(y)\}=0 \,\,;\,\,
\nn 
&\{\psi^\dagger_{f,\alpha}(x),\psi_{l,\beta}(y)\}= \delta(x-y)\, \delta_{f,l} \, \delta_{\alpha,\beta}\, ,
\end{align}
where $\alpha,\beta$ are spinor indices (left implicit in what follows). The electric field satisfies Gauss' law,  
$\partial_x E=g \sum_{f } \psi_{f}^\dagger \psi_{f}$;  it is not
a dynamical degree of freedom and can be explicitly integrated out. 

 To adapt this model to our problem of hyperon spin correlations, 
 we first make the simplifying assumption that since the ratio of the physical up and down quark masses is 
 $m_u/m_d\sim \mathcal{O}(1)$~\cite{ParticleDataGroup:2022pth},  their dynamics are indistinguishable. 
 Further,  since the  mass gap between  the heavy strange flavor and these light flavors is $m_s/(m_d+m_u)\sim \mathcal{O}(10)$, it is sufficient to  
 simply consider a light and a heavy flavor fermion in our study; we will discuss later how this isospin symmetry can be lifted in our model. 
 
 More importantly, the structure of the Lorentz group dictates that there is no notion of spin in 1+1-dimensions.
  We will show here that one can nevertheless construct an effective model of spin dynamics using the map [flavor ($f$) $\to$ (species ($s$), spin ($\sigma$))]:
 \begin{align}\label{eq:spin_labels}
1 \to (h,\ua)\,\,;\,\, 2 \to (h,\da) \,\,;\,\,
3 \to (l,\ua) \,\,;\,\, 4 \to (l,\da) \,,
\end{align}
where the $h,l$ indices denote the heavy and light fermions, with masses $m_h$ and $m_l\ll m_h$, while $\ua,\da$ correspond to their up or down spin state. Using this  double label $(s,\sigma)$, we have therefore mapped the four two-component spinors to two four-component spinors, and Eq.~\eqref{eq:com_rels} satisfies the  3+1-d QED anticommutation relations with $\delta_{l,f}\to \delta_{s,s'}\delta_{\sigma,\sigma'}$. Due to the large mass gap between fermionic species, the global flavor symmetry group of Eq.~(\ref{eq:H_Schwinger}) reduces from ${\rm SU}_{f}(4)\to {\rm SU}_h(2)\times {\rm SU}_l(2)$. 

An important difference from 3+1-d QED is the absence of magnetic fields in 1+1-d; there are therefore no Thomas precession or Larmor interaction terms. However we anticipate that such interactions would be suppressed for our problem of interest~\footnote{These terms in the light-heavy system are suppressed by the large invariant mass, as are spin-flip interactions of ultrarelativistic light quarks. In this \textit{adiabatic} approximation, spins live on a Bloch sphere with their directions corresponding to superpositions of distinct flavor states.}. The spin-dependent interactions of light and heavy flavors in the QCD string are therefore well-approximated by adding to Eq.~(\ref{eq:H_Schwinger}) the effective spin Hamiltonian
\begin{align}
H_{\rm spin} &= \int dx\,g_{ll}^0 \, \bar \psi_{l,\ua}\gamma^0 \psi_{l,\da}
	+g_{ll}^1 \, \bar \psi_{l,\ua}\psi_{l,\da} \nn
	&+ g_{lh}^0(\bar \psi_{h,\ua}\gamma^0 \psi_{l,\da} +\bar \psi_{h,\da}\gamma^0 \psi_{l,\ua}   ) \nn
	&+g_{lh}^1(\bar \psi_{h,\ua} \psi_{l,\da} +\bar \psi_{h,\da}\psi_{l,\ua} ) + {\rm h.c.} \, ,
\end{align}
where we used the mapping in Eq.~\eqref{eq:spin_labels}. The first two terms describe the spin-flip interactions between the light fermions with their respective coupling constants. The subsequent terms describe the spin interactions between light and heavy fermions. Direct heavy spin-flip interactions 
are assumed to be suppressed by their large mass~\footnote{The  smaller multiplicity of heavy relative to light quarks makes such interactions sub-dominant.}. The addition of $H_{\rm spin}$ reduces the  global symmetry of the $N_f=4$ Schwinger 
model from ${\rm SU}_h(2)\times {\rm SU}_l(2)\to  {\rm SU}_{\rm spin}(2) $ corresponding to the residual symmetry group of light and heavy quarks in the QCD string~\footnote{This only includes the minimal set of operators preserving ${\rm SU}(2)_{\rm spin}$, with operators such as $\bar \psi_{h,\ua} \psi_{l,\ua}$ disallowed.}. 
\begin{figure}
    \centering
    \includegraphics[width=.45\textwidth]{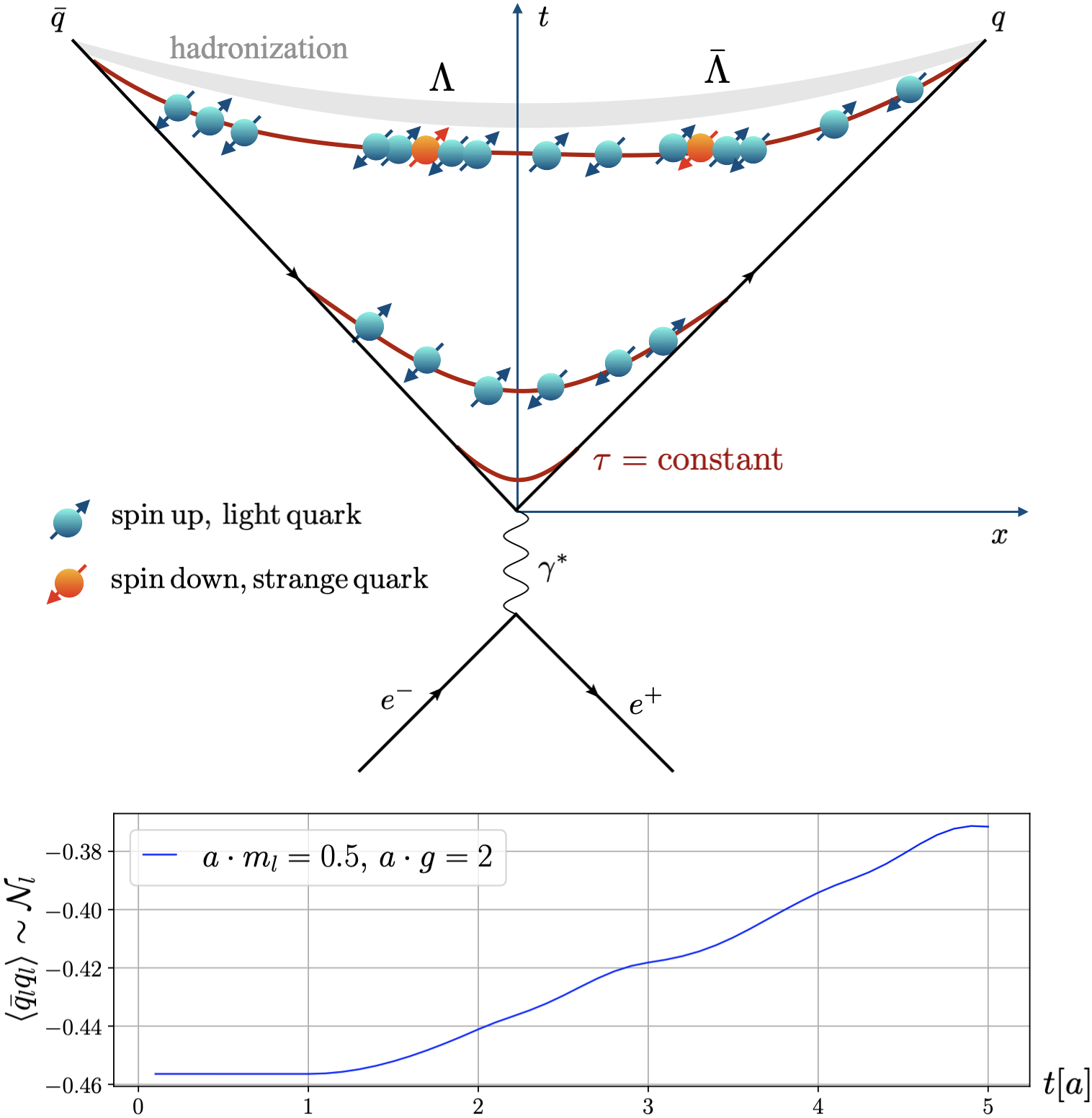}
    \caption{\textbf{Top:} Illustration of realtime evolution of 
    the confining string described by the model Hamiltonian $H=H_{\rm Schwinger}+ H_{\rm Spin}$. The string stretched between a $q\bar q$ pair moving along the forward light cone is increasingly populated by light quarks with each proper time $\tau$ slice. The heavy quark multiplicity is suppressed due to its larger mass and its  spin  is modified primarily due to multiparticle interactions.
    \textbf{Bottom:} Lattice computation in the $N_f=1$ massive Schwinger model for the light quark correlator $\langle \bar q_l q_l\rangle$ correlator. For  $m_l<g$, the increase of this observable is directly related to a larger multiplicity of light quarks $\mathcal{N}_l$ within the string as shown in the illustration. Details of figures in main text. 
    }
    \label{fig:phys_pic}
\end{figure}

The  model deformed Hamiltonian 
$H = \Hs + H_{\rm spin}$ can now be employed to explore the realtime dynamics of spin correlations between heavy fermions in the dynamical process $e^+e^-\rightarrow {q{\bar q}}\rightarrow \Lambda {\bar \Lambda} +X$. This is illustrated in Fig.~\ref{fig:phys_pic} (\textbf{top}). Recall that, inspired by the nonrelativistic quark model, we treat strange quark (antiquark) spin as a proxy for $\Lambda$ (${\bar \Lambda}$) spin in our model. 

The qualitative picture of this process is as follows. The injection of energy from the virtual photon forms the primordial 
$q{\bar q}$ pair.  Additional pair production taking place due to the Schwinger mechanism~\cite{grib1994vacuum,Schwinger:1951nm} is demonstrated by explicit computation using tensor network techniques (discussed below) for the $N_f=1$ case in Fig.~\ref{fig:phys_pic} (\textbf{bottom}) for an expanding QCD string. As suggested by Fig.~\ref{fig:phys_pic} (\textbf{top}), the string should be mostly populated by light quarks. The light and strange quarks carry spin and their self-interactions dynamically generate nontrivial correlations between the rarer heavy strange quark pairs, with the  multiparton quantum dynamics reflected in the $\Lambda \bar\Lambda$ spin correlations measured in hyperon weak decays. 

This complex dynamical evolution can be captured by extending the gauge 
sector in Eq.~\eqref{eq:H_Schwinger} to~\cite{Casher:1974vf,Honda:2021ovk,Nagano:2023uaq} 
\begin{align}\label{eq:E2_lattice_with_ext_field}
\int dx\, E^2(x)	\to  \int dx\,(E(x)+E_{\rm ext.}(x))^2 \, .
\end{align}
The external field $E_{\rm ext.}$ plays the role of the embedding string, see  Fig.~\ref{fig:phys_pic} (\textbf{top}), taking the form
\begin{align}
   E_{\rm ext.} = -|Q| \, \Theta(x^<(t) <x<x^>(t)) \, ,
\end{align} 
with $|Q|$ the absolute value of the external charges (the $q\bar q$ pair in  Fig.~\ref{fig:phys_pic} (\textbf{top})) generating the field and $x^{>(<)}=+(-)t$ referring to their right (left) dynamical spatial positions along the forward lightcone. For analogous detailed studies of realtime evolution of pair production, charge separation and screening for the $N_f=1$ Schwinger model, see  
~\cite{Hebenstreit:2013qxa,Hebenstreit:2013baa,Hebenstreit:2014rha,Lee:2023urk,Florio:2023dke}.

We will now explore in our $N_f=4$ framework the quantitative realization of the picture we have outlined. We discretize the Hamiltonian $H$ by employing staggered 
fermions~\cite{Susskind:1976jm,Banks:1975gq} on a lattice with $N$ sites and lattice spacing $a$. 
The staggered discretization of the dimensionful continuous two component spinor $\psi_f$ leads to the dimensionless single component spinor $\chi_{s,\sigma}(\tilde n)$ on the lattice; they are related as
\begin{align}\label{eq:help_1}
   \psi_f(x=2 \tilde na)  \to \frac{1}{\sqrt{a}}\begin{pmatrix}
		\chi_f(2 \tilde n)\\
		\chi_f(2  \tilde n-1)
	\end{pmatrix}\, .
\end{align}
 The lattice index $\tilde n=1,2,\cdots \tilde N= N/4$ labels the staggered sites; for each one of these, there are four computational lattice sites, labeled by $n=1,2,\cdots, N$, that follow the ordering implicit in Eq.~\eqref{eq:spin_labels}~\footnote{One therefore identifies $\chi(
\tilde n + (f-1)_{\mod 4}) = \chi_{s,\sigma}(\tilde n) $, using Eq.~\eqref{eq:spin_labels}.}. The single component spinor $\chi_{s,\sigma}(\tilde n)$ represents fermions on even sites and antifermions on odd sites and it satisfies the commutation relations
\begin{align}
	&\{\chi_{s,\sigma}(\tilde n),\chi_{s',\sigma'}(\tilde m)\}=0 \,,\nn
 &\{\chi_{s,\sigma}^\dagger(\tilde n),\chi_{s',\sigma'}(\tilde m)\}= \delta_{\tilde n,\tilde m}\delta_{s,s'}\delta_{\sigma,\sigma'} 	\, .
\end{align}
 Since  Gauss' law, as noted previously, dictates that the electric field is not dynamical, all dependence on gauge fields can be integrated out, resulting in a (nonlocal) expression for the electric field and $H_{\rm Schwinger}$ entirely in terms of $\chi$. Indeed, using open boundary conditions with $L(0)=0$, the electric field $L= E/g$ on the $n$th link is given by 
\begin{align}\label{eq:Electric_field_latt}
L(n)&=L( n-1)\nn 
&+ \sum_{f} \Bigg( \chi_{f}^\dagger(  n) \chi_{f}(n) -\frac{1}{2}(1-(-1)^{\tilde n})\Bigg)\, ,
\end{align}
where the sum over the flavors is not made fully explicit. As a result, using Eq.~\eqref{eq:help_1}, we can write the different lattice elements entering $H$; for the fermionic sector we have~\cite{Susskind:1976jm}:
\begin{align}
  &\int dx \,m_f \bar \psi_f(x) \psi_f(x)\to    \sum_{\tilde n=1}^{N} m_f (-1)^{\tilde n} \chi^\dagger_f( \tilde n)\chi_f(\tilde n) \, , \nn
&\int dx\, \bar  \psi_f(x) (-i \gamma^1 \partial_1 +g \gamma^1 A_1(x) ) \psi_f(x) \nn 
&\to -\frac{i}{2a} \sum_{\tilde n=1}^{\tilde N-1}    \chi_f^\dagger(\tilde n)  \chi_f(\tilde n+1) - {\rm h.c.} \, ,
 \end{align}
 where we have performed a residual gauge transformation~\cite{Hamer:1997dx} to obtain an explicitly fermionic theory. Note that here the index dependence on the flavor label is not made fully explicit. We provide the full dependence below in the spin Hamiltonian, which admits a similar treatment:
 \begin{align}
&H_{\rm spin } \to  \sum_{\tilde n=1}^{\tilde N} 	(g_{ll}^0+ g_{ll}^1(-1)^{\tilde n}) (\chi^\dagger_{l,\ua}(n)\chi_{l,\da}(n)) \nn 
&+	(g_{lh}^0+ g_{lh}^1(-1)^{\tilde n}) (\chi^\dagger_{h,\ua}(\tilde n)\chi_{l,\da}(\tilde n)+\chi^\dagger_{h,\da}(\tilde n)\chi_{l,\ua}(\tilde n)) \nn 
&+ {\rm h.c.}\, .
\end{align}
Finally, using Eq.~\eqref{eq:Electric_field_latt}, the pure gauge term in $H$ reduces to
 \begin{align}\label{eq:integrated_E2_lattice}
 \frac{1}{2} g^2 a \sum_{n=1}^{\tilde N-1} \Big[\sum_{f}\sum_{\tilde k=1}^{\tilde n} \chi^\dagger_{f}(\tilde k)\chi_f(\tilde k) -\frac{1}{2}(1-(-1)^{\tilde k}) \Big]^2\, .
 \end{align}

 With the goal of quantum simulating the model, we further map the lattice Hamiltonian to  a one-dimensional spin chain model via a multiflavor Jordan Wigner transform (JWt)~\cite{Steinhardt77,Banks:1975gq,Steinhardt77,Banuls:2016gid}.
For our particular case, it can be expressed as 
\begin{align}\label{eq:jwt}
	&\chi_{h,\uparrow}(\tilde n) =   S(\tilde n) \sigma^-_{h ,\uparrow}(\tilde n)\, ;\nn 
	&\chi_{h,\downarrow}(\tilde n) =  i S(\tilde n)\sigma^z_{h ,\uparrow}(\tilde n)  \sigma^-_{h ,\downarrow}(\tilde n) \, ;\nn 
	&\chi_{l,\uparrow}(\tilde n) =  -  S(\tilde n)\sigma^z_{h, \uparrow}(\tilde n) \sigma^z_{h ,\downarrow}( \tilde n)  \sigma^-_{l,\uparrow}(\tilde n) \, ;\nn 
	&\chi_{l,\downarrow}(\tilde n) =  -i S(\tilde n)\sigma^z_{h, \uparrow}( \tilde n) \sigma^z_{h ,\downarrow}(\tilde n)   \sigma^z_{l, \uparrow}(\tilde n) \sigma^-_{l ,\downarrow}( \tilde n) \, ;
\end{align}
where we introduced the string operator 
\begin{align}
  S(\tilde n)\equiv   \prod_{\tilde k<\tilde n}[\sigma^z_{h ,\uparrow}(\tilde k) \sigma^z_{h ,\downarrow}(\tilde k) \sigma^z_{l, \uparrow}(\tilde k) \sigma^z_{l,\downarrow}(\tilde k) ] \, ,
\end{align}
in terms of local Pauli operators $\sigma^i_{s,\sigma}(\tilde n)$ acting on site $\tilde n$ of the staggered lattice. The subscript indicates the flavor and spin indices and the superscript denotes the direction on the Bloch sphere. With these 
identifications, our model Hamiltonian can be mapped to a spin chain with the Hamiltonian~\cite{Itou:2023img,Banuls:2016gid}
\begin{widetext}
\begin{align}\label{eq:H_lattice}
	H_{\rm lat} &= 	\frac{1}{2} g^2 a \sum_{\tilde n=1}^{\tilde N-1} \Big[\frac{1}{2}\sum_{s,\sigma}\sum_{\tilde k=1}^{\tilde n} (\sigma^z_{s,\sigma}(\tilde k)+(-1)^{\tilde k}) \Big]^2 
  +\sum_{\tilde n=1}^{\tilde N}\sum_{s,\sigma} m_s  (-1)^{\tilde n} \left(\frac{1+\sigma^z_{s,\sigma}(\tilde n)}{2}\right) \nn
&+\frac{i}{2a}\sum_{n=1}^{N_f(\tilde N-1)} 	(\sigma^+(n) \sigma^z(n+1)\sigma^z(n+2)\sigma^z(n+3)\sigma^-(n+4) - \sigma^+(n+4) \sigma^z(n+3)\sigma^z(n+2)\sigma^z(n+1)\sigma^-(n)) \nn
	&+ \sum_{\tilde n=1}^{\tilde N} 	(-i)(g_{ll}^0+ g_{ll}^1(-1)^{ \tilde n}) (\sigma^+_{l,\ua}(\tilde n)\sigma^-_{l,\da}(\tilde n)-\sigma^+_{l,\da}(\tilde n)\sigma^-_{l,\ua}( \tilde n))  
	+	i(g_{lh}^0+ g_{lh}^1(-1)^{\tilde n}) ( \sigma^+_{h,\ua}(\tilde n)\sigma^z_{h,\da}(\tilde n) \sigma^z_{l,\ua}(\tilde n)\sigma^-_{l,\da}(\tilde n)\nn
 &-\sigma^+_{l,\da}(\tilde n)\sigma^z_{l,\ua}(\tilde n) \sigma^z_{h,\da}(\tilde n)\sigma^-_{h,\ua}( \tilde n)
	+\sigma_{l,\ua}^+(\tilde n)\sigma^-_{h,\da}(\tilde n)-\sigma_{h,\da}^+(\tilde n)\sigma^-_{l,\ua}(\tilde n)) \, ,
\end{align}
\end{widetext}
where $s={h,l}$ (or $1, 2$), $\sigma={\ua,\da}$ (or  $1, 2$), and the second sum is indexed with respect to the computational lattice. 

The explicit form of $H_{\rm lat}$ makes it  evident that the numerical simulation of real time evolution in this theory is highly nontrivial, especially when implementing
the picture detailed in Fig.~\ref{fig:phys_pic}. 
This problem therefore constitutes an ideal setting to use digital quantum computers to simulate real time processes in gauge theories that have direct contact with high energy physics phenomenology and are not likely easily simulated
using other methods. The circuit depth and system size necessary to see nontrivial effects are parametrically controlled by the total evolution time $t \sim L\gtrsim d_{\mathbf{c}}$; here $d_{\mathbf{c}}\propto \frac{1}{g}$ is the critical charge separation for nonlinear effects leading to string breaking to become important, assuming $g/m_l\ll 1$~\cite{Hebenstreit:2013baa}. Working in the gaugeless version of the theory, the total number of logical qubits would then scale as $n_{\rm qubits}\sim \mathcal{O}(1/(a g))$. For $a\cdot m_l=0.1$, one would thus need at a minimum $\mathcal{O}(4\times 10)$ logical qubits to fully capture the system, ignoring lattice and staggering effects. 

The number of basic quantum gates depends on the particular decomposition of the Hamiltonian, but its cost will be dominated by the nonlocal gauge term. Note that since the spin terms do not require extra qubits, or introduce more complex operators at the Hamiltonian level, the simulation complexity should grow parametrically as in the \textit{vanilla} $N_f=4$ Schwinger model. A detailed analysis of the complexity and system size growth can be found in ~\cite{Shaw:2020udc} for the $N_f=1$ theory; naively extrapolating to the $N_f=4$ case, one should expect a linear growth in the number of qubits, while the number of basic quantum gates should grow exponentially, due to the all-to-all form of the pure gauge term.

Due to the current limitations of quantum devices to perform sufficiently long simulations, we will study the time evolution and spatial correlations present in this model separately using two different approaches. To test the quantum dynamics of the time evolution of the QCD string, we will use the exact diagonalization (ED) method. This allows one to perform exact time evolution, albeit for small system sizes. To study  spatial and spin correlations between the light and heavy quarks, we will consider a longer spin chain which is limited by current resources to evolve only for short time intervals. For this case, we will therefore employ tensor network methods.

We first discuss the longer time evolution of the QCD string using ED methods. We employ a lattice of $N=24$ sites, and initially prepare the ground state $\ket{GS}$ of $H_{\rm lat}$. To simulate the effect of the external field $E_{\rm ext.}$, we construct a time-dependent Hamiltonian $H_{\rm ext.}(t)$ by rewriting the gauge term in Eq.~\eqref{eq:H_lattice} as
\begin{align}
    \frac{1}{2} g^2 a \sum_{\tilde n=1}^{\tilde N-1} \Big[\frac{1}{2}\sum_{s,\sigma}\sum_{\tilde k=1}^{\tilde n} (\sigma^z_{s,\sigma}(\tilde k)+(-1)^{\tilde k}) \notag \\- Q\, \Theta(-x(t)+1 \leq \tilde{k} - N/8 <x(t)) \Big]^2,
\end{align}
where $x(t) = t$ for discrete time steps $t = 1, 2 \cdots, N/8$. For instance, for $N = 24$, the external field extends from $3 \leq \tilde{k} < 4$ at $t = 1$, from $2 \leq \tilde{k} < 5$ at $t = 2$, etc.. Evolving $\ket{GS}$ under $H_{\rm ext.}(t)$ for these discrete time steps thus simulates a particle-antiparticle pair with charge $Q$ initialized at the center of the lattice and moving apart at speed $c = a/t$. Time evolution is computed using the Krylov subspace method, which involves approximating $H\ket{\psi}$ by restricting evolution to a smaller Hilbert space generated by the repeated action of $H$ on $\ket{\psi}$ \cite{Bochev2012}. Our flavor ordering on the lattice follows the convention in Eq.~\eqref{eq:jwt}.

\begin{figure*}
    \centering
    \includegraphics[]{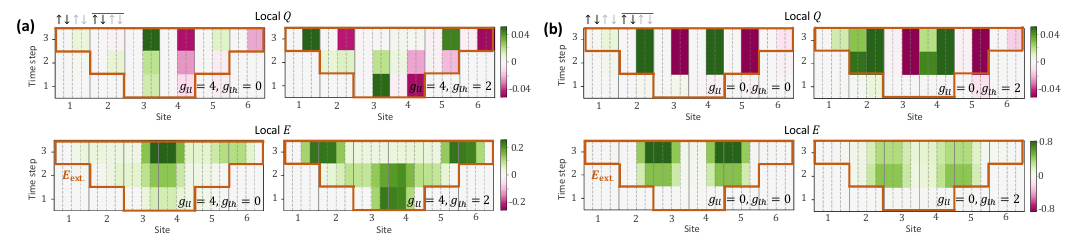}
    \caption{\textbf{Top (a, b):} Local charge density of dynamics under $H_{\rm ext.}(t)$ for the different parameters indicated in the text; ground state values are subtracted. The region over which the external field acts is outlined at each time step. Four flavors--- $h_\uparrow, h_\downarrow, l_\uparrow, l_\downarrow$--- are located at each of six staggered lattice sites, for a total of $N=24$ sites. Pink and green regions indicate particle-antiparticle pairs of light quarks in the string. \textbf{Bottom (a, b):} Local electric field strength of dynamics under $H_{\rm ext.}(t)$ for the parameters indicated in the text, with ground state values subtracted. Pair production induced by the external electric field results in a local field of opposite direction, thus leading to screening.}
    \label{fig:ed_tot}
\end{figure*}

At each time step, we compute the internal electric field $E_i$ and charge density $Q_i$ at each lattice site $i = s' + \sigma' + \tilde n$, where
\begin{align}
    \langle E_i \rangle & = \frac{g}{2} \sum_{s,\sigma}^{s',\sigma'}\sum_{\tilde k=1}^{\tilde n} \braket{\sigma^z_{s,\sigma}(\tilde k)+(-1)^{\tilde k}}, \notag\\
    \langle Q_i \rangle &= \braket{\chi_{s',\sigma'}^\dagger(\tilde n),\chi_{s,\sigma}(\tilde n)} = \frac{1}{2a}\braket{\sigma^z_{s',\sigma'}(\tilde n)+(-1)^{\tilde n}}, \notag
\end{align}
with the ground state values implicitly subtracted. Note that the internal electric field does not include the external field imposed by the $q\bar{q}$ pair flying along the light cone. We use a representative parameter set $a \cdot g = 2$, $a \cdot m_l = 0.5$,  $a \cdot m_h = 7.5$, $Q=1$, $g^{(0, 1)}_{ll}=0, 4$, and $g^{(0, 1)}_{lh}=0, 2$, with the numerical results in Fig.~\ref{fig:ed_tot}.  

As seen from the local $Q$, for all sets of parameters $g_{ll}=0, 4$, $g_{lh}=0, 2$, $l\bar{l}$ pairs of both spins are induced in the vacuum by the external field---this exemplifies the Schwinger pair production mechanism in our four-flavor model. These pairs result in a positive electric field within the light cone thus partially screening the negative external field $-Q$. Moreover when $a \cdot g_{ll} = 4$, $a \cdot g_{lh} = 0$, the lack of spin-flip interactions between heavy and light particles localizes pair production within light particles at the center of the string. In contrast, when $a \cdot g_{lh} = 4$, $a \cdot g_{lh} = 2$, the additional kinetic exchange between heavy and light spins allows pair production to be seen across all sites within the light cone of the expanding external charges. When $a \cdot g_{ll} = 0$, $a \cdot g_{lh} = 0$, the bare four-flavor Schwinger model with anisotropic masses is simulated. In this case, pair production occurs nearly uniformly across the center of the string. Only minor differences from the bare four-flavor model are visible when $a \cdot g_{ll} = 0$, $a \cdot g_{lh} = 2$, indicating that fast dynamics in the system are generally mediated by light-light quark interactions. Notably for all cases, pair production is seen predominantly in the light species; this phenomenon is to be expected because the Schwinger pair production formula tells us that heavier pairs should be exponentially suppressed relative to light ones.

\begin{figure}
  \centering
  \includegraphics[width=.45\textwidth]{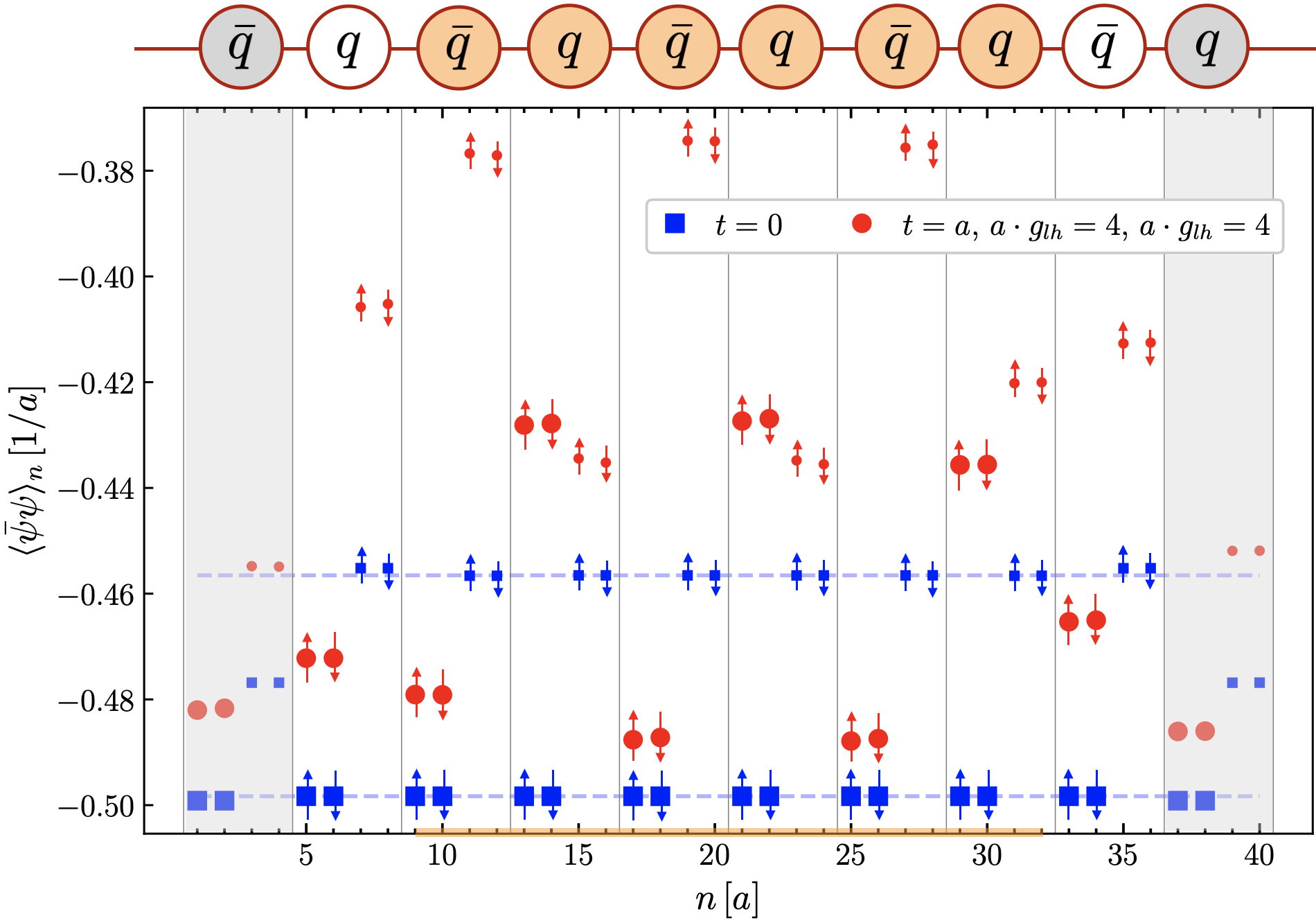}
  \caption{Lattice distribution of local one point function $\langle \bar \psi \psi\rangle_n$. At top is depicted the staggered construction alternating between fermions ($q$) and antifermions ($\bar q$). Each staggered site corresponds to four computational sites. Gray bands mark where  
   lattice artifacts 
   prevail. Gold band at the bottom (and  gold $q$/$\bar q$ symbols) indicates where the external electric is active. 
  Thick [thin] blue and red symbols represent respectively heavy [light] fermion sites for ground ($t=0$) and excited ($t=a$) states respectively. Particle spins are denoted by $\uparrow \downarrow$. Dashed lines denote the ground state condensate for each particle species.}
  \label{fig:condensate}
\end{figure}

Having explored the time evolution of the system as illustrated in Fig.~\ref{fig:phys_pic}, we shall now consider the build up of spatial correlations after a short evolution. Taking into account the well-known limitations of tensor networks in performing real time evolution, we will restrict ourselves to a simpler yet highly informative setup. Using a $N=40$ site lattice, we first prepare the ground state of the system for $g_{ll}^0=g_{ll}^1=g_{lh}^0=g_{lh}^0=0$, in other words, the ground state of the four-flavor massive Schwinger model at finite coupling~\footnote{One could also prepare the ground state at finite couplings using the DMRG algorithm. However for the lattice size and parameter range studied, one observes a slower convergence with a larger $O(10)$ maximal bond dimension compared to the pure Schwinger model case.}. We adopt a matrix product state (MPS) architecture using the density matrix renormalization group (DMRG) algorithm~\cite{PhysRevLett.69.2863,PhysRevB.48.10345}. We then time evolve the system under the full Hamiltonian with finite spin couplings, inserting a static external electric in the middle of the lattice (between lattice sites 9 and 32) as shown in  Fig.~\ref{fig:condensate}. The evolution time is chosen to be $t=a$~\footnote{For the parameters studied, longer evolution times require long computational times. Nevertheless, the time interval studied is sufficient to observe significant build up of the particle condensate and spin correlations.} and is performed using the time-dependent variational principle (TDVP) algorithm~\cite{Haegeman:2011zz,PhysRevB.94.165116}. 

Observables are computed on the time evolved state in the region spanned by the external field; note that its spatial extent is chosen to minimize boundary effects due to open boundary conditions and to maximize the number of usable lattice points. All simulations are performed using the tensor network package \texttt{ITensor}~\cite{itensor} in \texttt{Julia} for the parameter set $a\cdot g=2$, $a\cdot m_l=0.5$ and $a\cdot m_h=7.5$. The bare fermions masses are chosen such that their ratio is of the order of $m_s/(m_u+m_d)$ and they include the improvement term~\footnote{ For even $N_f$ and $m_f=0$, the lattice theory has a shift symmetry corresponding to a discrete chiral symmetry in the continuum~\cite{Dempsey:2022nys}. Its effect is however not visible in our massive $N_f=4$ case for the parameter range studied.} discussed in~\cite{Dempsey:2022nys,Dempsey:2023gib}.

To illustrate the key features of our model, we begin by computing the one point fermion correlator $\langle \bar \psi \psi\rangle_n$; the subscript denotes  the expectation value at site $n$. As shown in Fig.~\ref{fig:condensate}, the ground state is characterized by two condensates (in blue) for each of the two fermion species. The light fermions have a larger value for the condensate. After time evolution (in red), light fermion pairs are easily excited in the region where the electric field is activated (denoted by the gold $q$/$\bar q$ labels), and the heavy and light condensates disappear. 
For both species, the points are grouped in doublets; this reflects the residual ${\rm SU}_{\rm spin}(2)$ symmetry of the model.

To study spin correlations of the heavy (strange) quarks, we introduce the fermion correlator
\begin{align}\label{eq:C_corr}
        C_{\sigma,\sigma '}(\Delta ) \equiv  \frac{\sum\limits_{\tilde n,\tilde m} \langle \hat c_{\sigma}(\tilde n) \hat c_{\sigma '}(\tilde m)\rangle_{\rm conn.}   \, \delta_{ \Delta, \tilde n -\tilde  m}}{\sum\limits_{\tilde n,\tilde m}    \delta_{ \Delta, \tilde n -\tilde  m}}\, ,
\end{align}
where $\Delta=1,2,3,\cdots$ is the spatial lattice separation, $\tilde n$, $\tilde m$ are chosen such that they correlate fermions with antifermions, with the operator  
\begin{align}
\hat c_\sigma (\tilde n) = \bar \psi_{h,\sigma} \psi_{h,\sigma}  - \langle  \bar \psi_{h,\sigma} \psi_{h,\sigma}\rangle_{\rm \ket{\psi}_{t=0}} \, ,
\end{align}
evaluated at the staggered site $\tilde n$. We subtract the ground state expectation value to minimize the dependence on the initial state; the connected correlator eliminates classical correlations.  

\begin{figure}[htbp]
  \centering
    \includegraphics[width=.38\textwidth]{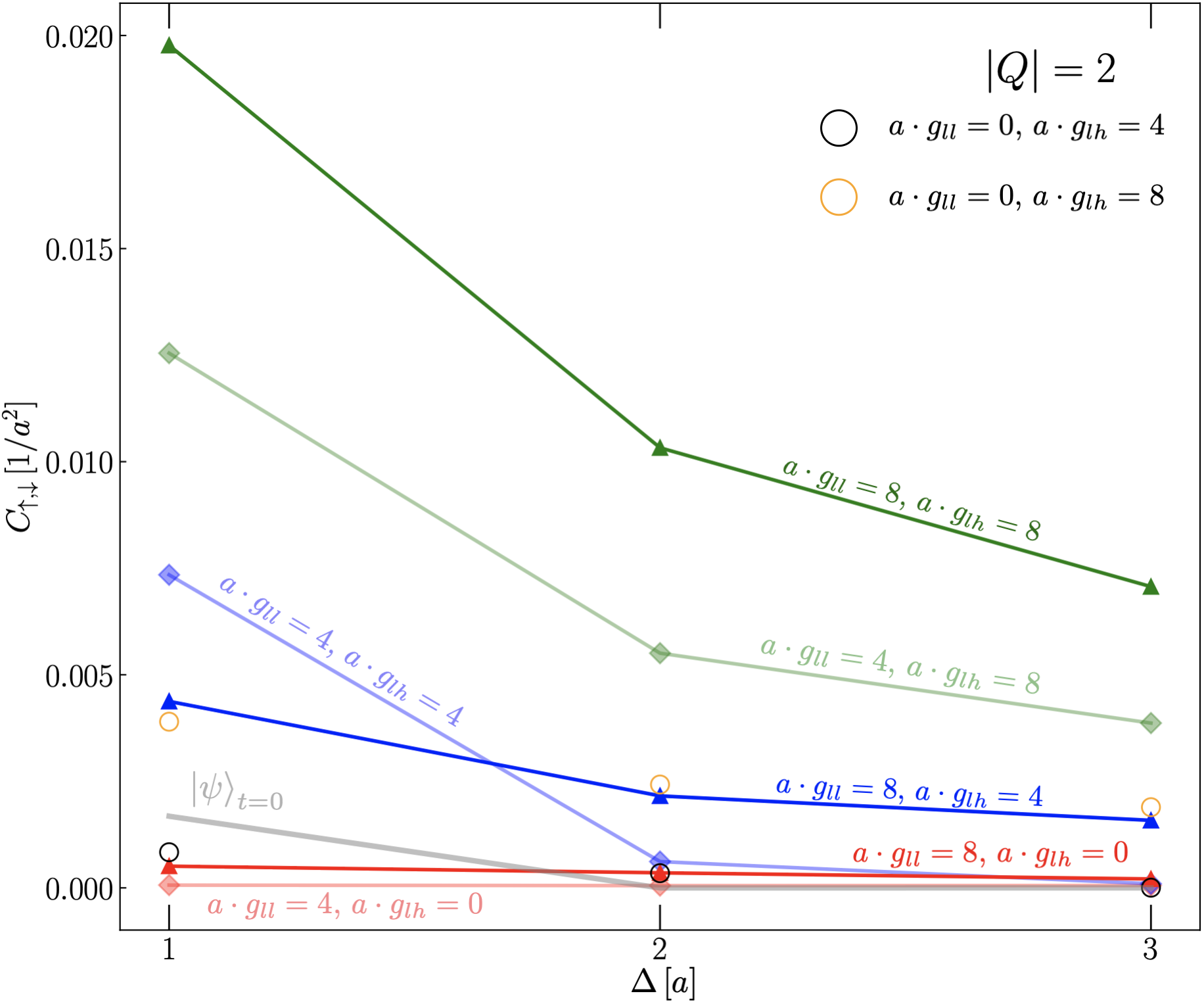}
  \caption{
  Heavy fermion correlator $C_{\ua,\da}$ as a function of the lattice separation $\Delta$. The electric field is generated with $|Q|=2$. The different colors depict the parameter sets used; the curve corresponding to the initial state is shown in gray.}
  \label{fig:correlator}
\end{figure}

In Fig.~\ref{fig:correlator} we show  numerical results for the $C_{\ua,\da}$ correlator. For the initial state (gray band), the correlator takes on small values and vanishes for large lattice separations. The same trend also occurs both for $g_{lh}=0$ (red and pink curves) corresponding to no spin-flips between the two fermions species, and for small, finite $g_{lh}$ at $g_{ll}=0$ (black and gold circles). The lack of large correlations between heavy fermions is expected when $g_{lh}=0$: in the absence of direct spin interactions between the heavy and light particles, there is no efficient mechanism to generate nontrivial correlations between heavy species. The fact that correlations are also weak at non-zero $g_{lh}$ but zero $g_{ll}$ indicates that the system prefers to generate correlations between heavy fermions indirectly through spin exchanges via the light degrees of freedom. This is consistent with what one would anticipate in QCD (see Fig.~\ref{fig:phys_pic}), where strange quark spin correlations are highly sensitive to the lighter degrees of freedom present in the string. Our conclusion is further supported by the other curves in Fig.~\ref{fig:correlator}. As $g_{lh}$  increases, generating stronger interactions between light and heavy fermions, short and long range correlations build up. Calculations performed for $C_{\ua,\da}$ exhibit quantitatively similar results. This is to be expected since our initial state is not prepared in a particular heavy fermion spin state. 


\begin{figure}[htbp]
  \centering
  \includegraphics[width=.45\textwidth]{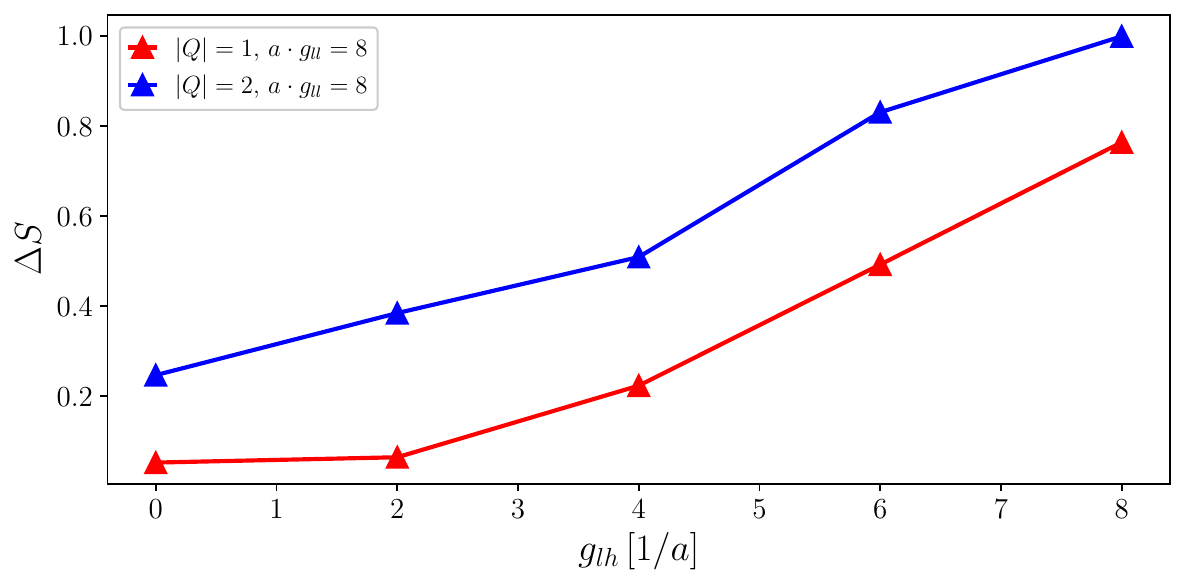}
  \caption{Half-chain entropy as a function of the $g_{lh}$ coupling.}
  \label{fig:entropy}
\end{figure}

Finally in Fig.~\ref{fig:entropy}, we plot the half-chain entanglement entropy, which can be easily extracted from the MPS. Since the initial state is not empty, we subtract the entropy present in the initial state: $\Delta S = S-S_{\ket{\psi}_{t=0}}$. As expected, larger values of the electric field result in a larger entanglement entropy, while $\Delta S$ also increases as a functions of $g_{lh}$. These results are in agreement with the behavior seen in Fig.~\ref{fig:correlator}, with larger correlations being reflected in larger $\Delta S$. However, numerical simulations with other parameter sets show that the evolution of the entropy as a function of $g_{lh}$ can be nonmonotonic. This is not surprising since the connection between $\Delta S$ and  correlation functions of the form of $C_{\sigma,\sigma'}$ is in general nontrivial~\cite{Gong:2021bcp};  we leave a more in-depth study of the time dependence of $\Delta S$ to future work.

In summary, ongoing studies of $\Lambda$ hyperon spin correlations at colliders have the potential to offer unique insight into quantum features of fragmentation and hadronization in QCD strings. To model the underlying dynamics, we constructed and explored the simplest QFT that captures fundamental aspects of the formation and many-body dynamics of heavy and light flavors in a QCD string. Though relatively simple compared to QCD, this spin model has an extremely rich phase structure worthy of study in its own right, and it shares many of its qualitative features such as a confining potential, chiral symmetry breaking, etc..  As a result, the proposed setup offers a quantitative approach to study spin correlations in a QFT with a qualitative connection to QCD. This is in contrast to state-of-the-art hadronization models such as the Lund string picture. As noted, such 1+1-d qualitative models of nonperturbative aspects of QCD do not fully capture the intrinsic quantum dynamics of 1+1-d QFT analogs of QCD, and require extensive tuning to be compared to collider data. 

Our 1+1-d QFT framework can be brought in closer analogy to QCD in several ways. The isospin symmetry used in the light fermion sector can be easily lifted by introducing two additional flavors. Such a transformation allows one to distinguish between up and down quarks; as a result, one can directly measure spin correlations of hadrons instead of spin correlations of heavy fermions. These hadrons would of course be $U(1)$ hadrons but the generalization to other gauge groups is straightforward. However on a technical level the inclusion of hadrons into our model would lead to a significant increase in the simulation complexity. Another refinement of our model would be to augment the spin term $H_{\rm spin}$ beyond the simple form taken in this paper. For example, quartic operators can be considered; these allow for direct spin exchanges between heavy fermions mediated by the light degrees of freedom.

Further progress is at present hindered on several fronts that deserve further attention. While our tensor network study can potentially be extended to larger lattices by further numerical optimization, time evolution will still pose a challenge. Exact diagonalization  
can circumvent the latter but we found its application has to be restricted at present to small lattices~\cite{fut_paper}. Quantum computers have the potential to overcome both these constraints thereby demonstrating quantitative progress in a problem of fundamental interest in collider physics. 

From the modeling perspective, one can further simplify the problem by integrating over the light degrees of freedom since the desired measurements are of heavy fermion spin correlations. Such a program can in principle be accomplished in the limit where the light fermion density is large and the heavy degrees of freedom only interact with a time averaged fluctuation of these fields. In addition, classical-statistical techniques could also be employed to study the time evolution of the system, see e.g.~\cite{Hebenstreit:2013qxa,Hebenstreit:2013baa,Hebenstreit:2014rha}. Indeed the present setup offers a useful testing ground to probe where such classical methods capture the full evolution; doing so allows one to isolate truly quantum effects. We hope our manuscript spurs further developments in these directions.

\smallskip

\noindent{\bf Acknowledgments} \\
J.B and R.V are supported by the U.S. Department of Energy under contract DE-SC0012704.
This material is based upon work supported by the U.S. Department of Energy, Office of Science, National Quantum Information Science Research Centers, Co-design Center for Quantum Advantage (C2QA) under contract number DE-SC0012704. W.G is supported by the Center for Theoretical Physics at the Massachusetts Institute of Technology, the Paul and Daisy Soros Fellowship for New Americans, and the Hertz Foundation Fellowship. W.G would also like to thank the BNL Nuclear Theory Group and C2QA for their hospitality and support during the completion of this work. R.V.'s work on this topic was supported in part by the Simons Foundation under Award number 994318.
He would like to thank Ross Dempsey and Igor Klebanov for valuable discussions. We are also grateful to Adrien Florio and Andreas Weichselbaum for clarifications of their work.

\bibliographystyle{apsrev4-1}
\bibliography{references.bib}

\end{document}